\documentclass[12pt]{iopart}
\usepackage{cite}
\usepackage{graphicx}
\usepackage[colorlinks, citecolor = blue, urlcolor = blue]{hyperref}
\begin{document}

\title[Gr\"uneisen rule in cubic lanthanum cage systems]{Gr\"uneisen rule in cubic rare-earth cage systems : the examples of LaB$_6$ and LaPt$_{4}$Ge$_{12}$}

\author{Mehdi Amara, Christine Opagiste}
\address{Univ. Grenoble Alpes, CNRS, Grenoble INP\footnote{Institute of Engineering Univ. Grenoble Alpes.}, Institut N\'eel, 38000 Grenoble, France}
\author{Natalya Yu. Shitsevalova}
\address{Frantsevich Institute for Problems of Materials Science, National Academy of Sciences of Ukraine, Kyiv, 03680 Ukraine}
\ead{mehdi.amara@neel.cnrs.fr}
\vspace{10pt}
\begin{indented}
\item[]
\end{indented}

\begin{abstract}
Specific heat and thermal expansion properties are investigated in two non-magnetic rare-earth cage compounds, LaB$_6$ and LaPt$_{4}$Ge$_{12}$, which represent extremes in guest-to-cage mass ratio. Using simplified phonons dispersions for the two lowest branches, a theoretical framework is proposed for the low temperature thermodynamic analysis of cage compounds. Within the quasi-harmonic approximation, the Gr\"uneisen rule is found to break down even at low temperatures. However, under the influence of the flattened branches, it should be approximatively restored at intermediates temperatures. The model accurately describes LaB$_6$ specific heat below 50 K. In  the LaPt$_{4}$Ge$_{12}$ case, the description is rapidly inadequate with increasing the temperature, which points to the interference of additional low frequency phonon branches. Subsequently, thermal expansion measurements are used to investigate the Gr\"uneisen rule in these two compounds. As predicted, there appears to be distincts Gr\"uneisen regimes at low temperature.  This study will help distinguish between phonon and magnetic contributions to the thermal expansion in the RB$_6$ and RPt$_{4}$Ge$_{12}$ series.   

\end{abstract}

%
%
%
%
%

\section{Introduction}
Cage compounds are characterised by a crystallographic structure in which tightly bonded atoms surround a larger element\cite{Keppens1998}. Inside this 'cage,' the guest atom has significant freedom of movement, a phenomenon often referred to as 'rattling'. These supposedly local modes were expected to reduce the thermal conductivity, which motivated the search, among cage systems, of materials with improved thermoelectric efficiency \cite{Sales1996}. Some cage compounds can accommodate a magnetic ion inside the cage, which grants the system additional displacement degrees of freedom, alongside those of the ion spin and orbit. How these degrees of freedom cooperate, resulting in specific static and dynamic properties, as well as new kind of ordering processes, remains a largely unexplored field of investigation.\newline
As demonstrated by the analysis of the antiferromagnetism in the rare-earth hexaborides series (RB$_6$), displacements inside the cage may play an essential role in the magnetically ordered states. In GdB$_6$, TbB$_6$ and PrB$_6$ \cite{Amara2005, Amara2010, Walker2009}, X-ray diffraction experiments show that, concomitantly with the antiferromagnetic order, static displacements waves of the rare-earth guest are stabilized. Indeed, the rare-earth displacements can minimize the exchange energy \cite{Amara2005} for specific magnetic wave vectors, among which $\left\langle \frac{1}{4}\;\frac{1}{4}\;\frac{1}{2} \right\rangle$, recurrent in the RB$_6$ series. In rare-earth hexaborides, the cage context is then active in the definition of the magnetic wave vector and, moreover, determines the first-order nature of the magnetic transition. \newline
Another distinctive feature of the cage environment arises from the crystalline electric field (CEF). In the paramagnetic state, the static energy scheme of the 4$f$ electrons is shaped by the CEF, reflecting the anisotropy of the lanthanide ion environment. This anisotropy should be consistent with the point symmetry at the rare-earth site. However, inside the cage, this symmetry applies only on average, as the rare-earth ion can depart significantly from its central position. Consequently, CEF schemes that would exhibit, at the cage center, an extra degeneracy relative to the Kramers minimum, acquire a dynamic width \cite{Amara2019}. At temperatures close or below the energy width of a non-Kramers CEF ground-state, one can predict a reduction in the magnetic entropy and of the ionic susceptibilities, as well as a broader distribution of the rare-earth ion within the cage.\newline
This change in the distribution should affect the system's volume, resulting in a thermal expansion anomaly. To investigate these effects, one has to first determine the thermal expansion in absence of magnetic contributions. In rare-earth compounds, the non-magnetic contributions are typically evaluated by measuring a non-magnetic element in the series, such as the lanthanum (better suited to the light rare-earth series) or yttrium (for the heavier elements) based ones. Some modelling is then required to adapt the non-magnetic background to a particular magnetic element. For instance, in the case of specific heat measurements, one can take advantage of the Debye model for, via a mass correction, adjusting the phonons contribution to the magnetic elements. In the case of rare-earth hexaborides, the Debye approach cannot cope with the specificity of the cage compound, in particular the low energy, flattened, phonon branches.  In ref. \cite{Amara2020}, we proposed a two-frequency model, that accounts for the two lowest phonon branches, for deriving the non-magnetic background of CeB$_6$ from the LaB$_6$ specific heat data. \newline 
In the present work, we extend this model within the quasi-harmonic approximation to describe the phonon contribution to thermal expansion. This approach allows us to define distinct temperature ranges in which an adapted form of the Gr\"uneisen rule\cite{Grueneisen1908}, specific to cage compounds, may apply. Within this framework, specific heat and thermal expansion measurements of LaB$_6$ are then analysed.\newline  
In addition to this investigation of a non-magnetic rare-earth hexaboride, the same approach is applied to the non-magnetic filled skutterudite compound LaPt$_4$Ge$_{12}$\cite{Gumeniuk2008, Gumeniuk2011}. This is a distinct type of rare-earth cage compound, crystallizing in the body-centered cubic LaFe$_4$P$_{12}$-type structure \cite{Jeitschko1977}, unlike the primitive cubic lattice of rare-earth hexaborides. The most significant difference from LaB$_6$, that affects the dynamic properties, is the guest-to-cage mass ratio, characteristic of a heavy guest in LaB$_6$ against a heavy cage for LaPt$_4$Ge$_{12}$. Rare-earth filled skutterudites exhibit a range of intriguing properties, such as the heavy-fermion and superconducting characteristics of PrOs$_4$Sb$_{12}$ \cite{Bauer2002}, the metal-insulator transition in PrRu$_4$P$_{12}$\cite{Iwasa2005} and the non-magnetic ordering of PrFe$_4$P$_{12}$ \cite{Keller2001}. The focus is here restricted to the specific heat and thermal expansion properties of LaPt$_4$Ge$_{12}$, which are confronted with the two-frequency phonons model and discussed in relation to the LaB$_6$ cage compound paradigm.\newline

\section{Quasi harmonic model of the thermal expansion in cage systems}
\subsection{Phonons for a cubic lattice of cages}
Describing phonons in a lattice of cages using the Debye approximation has limitations due to the influence of quasi-local modes. On the other hand, the purely local Einstein model does not accurately fit the specific heat data. These difficulties inspired a hybrid approach that combines both models \cite{Mandrus2001}. In ref. \cite{Amara2020} we proposed a more realistic description, restricted to the two lowest phonon branches, that takes advantage of simplified dispersion relations. This stems from the observation in LaB$_6$ of similar dispersion curves for different polarizations and along different reciprocal space directions \cite{Smith1985}. Similar traits are observed in filled skuterrudites \cite{Lee2006,Koza2013}. This model was developed for the primitive cubic lattice of the RB$_6$ series and needs to be extended to the body centered cubic lattice of the filled skuterrudites RPt$_{4}$Ge$_{12}$. In both cases, the crystal is envisaged as a lattice of rigid, but elastically coupled cages of mass $M$, each enclosing a guest of mass $m$. In the cubic context, the harmonic approximation implies an isotropic potential well, with a natural frequency ${\omega_0}$ for the guest vibration in its cage. Considering a family of lattice planes, with spacing $d$, perpendicular to a unitary $\mathbf{u}$ vector, the classical equations of motion for a vibration mode with wave vector $\mathbf{q} = q \;\mathbf{u} =\left[ q_x , q_y , q_z \right]$ (the reciprocal components are dimensionless, the indices $x$, $y$ and $z$ referring to the cubic axes of the reciprocal lattice),  polarization $s$ and frequency $\omega_{s}(\mathbf{q})$ yield the relation:
\begin{equation}
\label{EqDisp1}
\cos (\frac{2 \pi}{a}  q d) = 1 - 2 (1 + \frac{m}{M} \frac{{{\omega _0}^2{\kern 1pt} }}{{{\omega _0}^2 - {\kern 1pt} {{\omega_{s}(\mathbf{q})}^2}}})\frac{{{\omega_{s}(\mathbf{q})^2}}}{{{\Omega_{s}(B)}^2}}
\end{equation}
where $a$ is the cubic lattice parameter,  $\Omega_{s}(B)$ is, for the $s$ polarization, the acoustic frequency for a lattice of empty cages (i.e. for $m=0$) at the border zone point $B$ in the $\mathbf{u}$ direction. The spacing $d$ can be related to the reciprocal lattice vector $\mathbf{\Gamma B}$, where $\Gamma$ is the origin : $ d = a/ 2| \mathbf{\Gamma B} |$. It is convenient to introduce reduced frequencies as proportions of the guest frequency ${\omega_0}$: 
\[
 y_s(\mathbf{q})=\frac{\omega_{s}(\mathbf{q})}{\omega_0} \textrm{ and } Y_s(B)=\frac{\Omega_{s}(B)}{\omega_0}
\]
Equation (\ref{EqDisp1}) then reads as:
\begin{equation}
\label{EqDisp2}
\cos (\pi \frac{q }{| \mathbf{\Gamma B} |} ) = 1 - \frac{2}{{Y_s(B)}^2} (1 + \frac{\rho}{1 - {y_{s}(\mathbf{q})}^2}) {y_{s}(\mathbf{q})}^2
\end{equation}
where $\rho = m/M$ is the guest/cage mass ratio. This form allows to immediately identify the frequencies at the origin of the Brillouin zone, where $\cos (\pi \frac{q }{| \mathbf{\Gamma B} |} )=1$. In addition to the zero frequency of the acoustic branch, one gets the optical frequency at the origin:
\[
 y_s^o(\mathbf{0})=\sqrt{1+\rho} \textrm{ or }  \omega_s^o(\mathbf{0})=\omega_0 \sqrt{1+\rho}
\]
This mode is the closest to the rattler picture and, for a guest in an isotropic harmonic well, is indifferent to the polarization.
Along $\mathbf{\Gamma B}$, from equation (\ref{EqDisp2}) and introducing the function:
\begin{equation}
\label{funcdisp}
c_s(\mathbf{q}) = {Y_s(B)}^2 \frac{1- \cos (\pi \frac{q }{| \mathbf{\Gamma B} |} ) }{2}
\end{equation},
one can define the two lowest phonon branches for an $s$ polarization :
\begin{itemize}
\item [-] Acoustic :
\begin{equation}
\label{acoust}
 y_s^a(\mathbf{q})= \sqrt{\frac{1+\rho+c_s(\mathbf{q})-\sqrt{\left(1+\rho+c_s(\mathbf{q})\right)^2-4 c_s(\mathbf{q})}}{2}}
\end{equation}
\item [-] Optic :
\begin{equation}
\label{optic}
 y_s^o(\mathbf{q})= \sqrt{\frac{1+\rho+c_s(\mathbf{q})+\sqrt{\left(1+\rho+c_s(\mathbf{q})\right)^2-4 c_s(\mathbf{q})}}{2}}
\end{equation}
\end{itemize}

\begin{figure}
\begin{center}
\includegraphics[width=12cm]{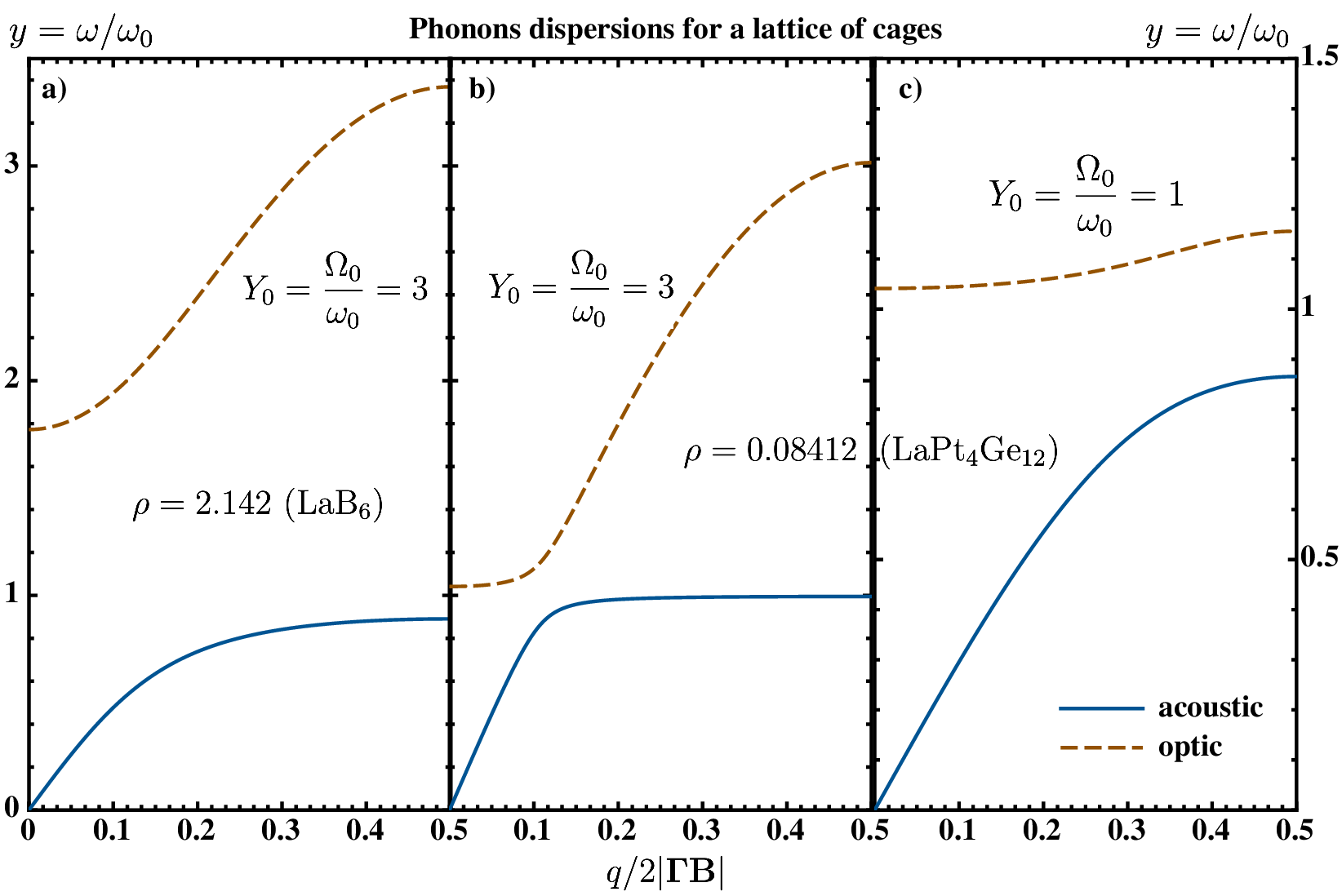}
\caption{\label{disp} Three examples of generic dispersion curves for the lowest branches of a lattice of cages, showing the reduced frequency $y = \omega /\omega_{0}$ as a function of the wave vector along a first Brillouin zone segment $\mathbf{\Gamma B}$. $\mathbf{a)}$ Heavy guest ($\rho >1$), hard lattice ($Y_0 = 3$) case, here for the $\rho$ mass ratio of LaB$_6$. $\mathbf{b)}$ Light guest, hard lattice case for the mass ratio of LaPt$_{4}$Ge$_{12}$, keeping $Y_0 = 3$.  $\mathbf{c)}$ Light guest, softer lattice ($Y_0 = 1$), for the mass ratio of LaPt$_{4}$Ge$_{12}$, using a different vertical scale.}
\end{center}
\end{figure}

These two branches materialize the avoided crossing between the flat mode at $\omega_{0}$ and the acoustic branch of a lattice of empty cages with upper frequency $\Omega_{0}$.  This results in features characteristic of a cage system, as illustrated by the examples of figure \ref{disp}:\newline
- an acoustic branch that flattens close to $\omega_{0}$, with an associated peak in the density of states. This effect is more pronounced for a light guest (small $\rho$) and hard lattice of cages, as in the case $\mathbf{b)}$ of figure \ref{disp}.  \newline
- the optical branch that starts at ${\omega^o_{s}(\mathbf{0})}= \omega_{0}\sqrt{1+\rho}$, separated from the acoustic one by a gap that increases with the $\rho$ inertia ratio (see the differences in the examples of figure \ref{disp}).\newline

In the model used, these traits are common to all wave vector directions and phonon polarizations, with quantitative differences arising only from variations in the $Y_s(B)$ value. This allows a further simplification of the model, wherein the value of $Y_s(B)$, specific to a zone border point and polarization, is replaced by a unique, averaged $Y_0$ value. Accordingly, the function $c_{s}(\mathbf{q})$ is replaced by $c_0(\mathbf{q})$, independent of the polarization. Phonon related properties, arising from a sum over all modes inside the first Brillouin zone, then depend on only two frequency parameters: $\omega_{0}$  and $\Omega_{0} = \omega_{0}\; Y_0 $. In some instances, the rattler frequency can be independently determined from spectroscopic techniques (such as neutron or infrared spectroscopy; see Ref. \cite{Smith1985} for the LaB$_6$ example), which leaves $\Omega_{0}$ as the only parameter to be adjusted. \\ 

\subsection{Calculation over the first Brillouin zone}

\begin{figure}
\begin{center}
\includegraphics[width=12cm]{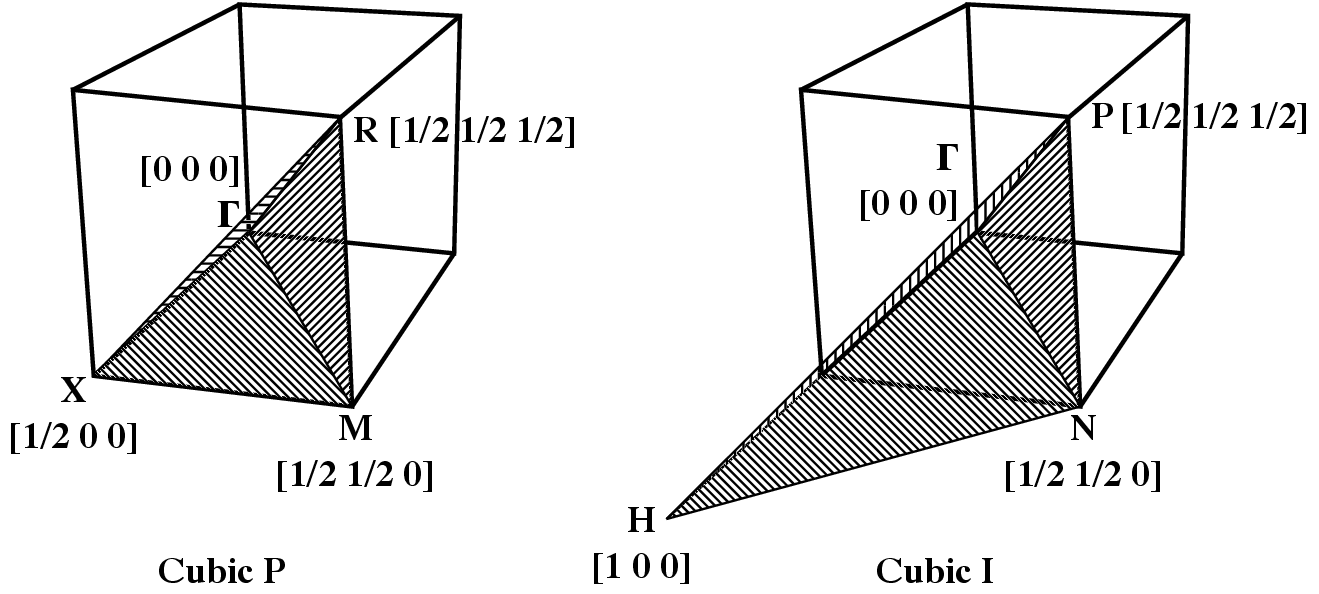}
\caption{\label{CubicBZ} Representative tetrahedra of the first Brillouin zone: left for a primitive cubic lattice (RB$_6$ case), right for a centered cubic lattice (filled skuterrudites case). Integrals over the first BZ volume can be approximated by a discrete sum of evenly distributed samples in these volumes.}
\end{center}
\end{figure}

Using the simplified dispersion functions, computing the specific heat, for instance, requires proper sampling of the Brillouin zone. Thanks to the cubic symmetry, one can restrict to samples evenly distributed within a representative volume of the first Brillouin zone. Figure \ref{CubicBZ} shows the tetrahedra enclosing such volumes in the case of a primitive cubic lattice (left, cubic P, that applies for LaB$_6$) and in the case of a centered cubic lattice (right, cubic I, for LaPt$_{4}$Ge$_{12}$). The border of the zone is the triangle XRM for the P case, respectively HPN for the I case. At points belonging to these triangles, the argument of the cosine $\cos (\pi \frac{q }{| \mathbf{\Gamma B} |} )$ in equations (\ref{EqDisp2}) and (\ref{funcdisp}) has to reach $\pi$. As these triangles are, respectively, perpendicular to $\mathbf{\Gamma X}=\left[ \frac{1}{2}, 0 , 0 \right]$ and to $\mathbf{\Gamma N}=\left[ \frac{1}{2}, \frac{1}{2} , 0 \right]$, for a wave vector $\mathbf{q}=\left[ q_x , q_y , q_z \right]$ inside the tetrahedra, one can rewrite the function $c_0(\mathbf{q})$, that replaces $c_s(\mathbf{q})$ in equations (\ref{acoust}) and (\ref{optic}), as:
\begin{itemize}
\item [-] Cubic P : \[c_0(\mathbf{q}) = \frac{{Y_0}^2}{2} \left[1- \cos (\pi \frac{\mathbf{q}\cdot\mathbf{\Gamma X}}{{\Gamma X}^2} )\right] = \frac{{Y_0}^2}{2} \left[1- \cos (2 \pi q_x )\right] \]
\item [-] Cubic I : \[c_0(\mathbf{q}) = \frac{{Y_0}^2}{2} \left[1- \cos (\pi \frac{\mathbf{q}\cdot\mathbf{\Gamma N}}{{\Gamma N}^2} )\right] = \frac{{Y_0}^2}{2} \left[1- \cos [ \pi (q_x+q_y) ]\right] \]
\end{itemize}
Once defined a collection of $N$ samples in the representative tetrahedron, these expressions allow for an easy implementation of the numerical calculation. For instance, a value for the phonons specific heat per unit volume is computed as:

\begin{equation}
\label{SpecHeat}
C_{p h}(T)=\frac{3 k_{\mathrm{B}}}{V}\sum_{i=1}^N  \sum_{\mu =(a,o)} \left(\frac{\hbar \omega_{\mu}(\mathbf{q}_i)}{k_{\mathrm{B}} T}\right)^2 \frac{e^{-\frac{\hbar \omega_{\mu}(\mathbf{q}_i)}{k_{\mathrm{B}} T}}}{\left(1-e^{-\frac{\hbar \omega_{\mu}(\mathbf{q}_i)}{k_{\mathrm{B}} T}}\right)^2} 
\end{equation}

where $V$ is the real space volume in correspondance with the $N$ reciprocal sample vectors. The factor 3 accounts for the 3 polarizations and the  index $\mu$ in the second summation refers to the phonons branches: the acoustic one ($\omega_{a}$) and the lowest optical ($\omega_{o}$) one. The $\omega_{\mu}(\mathbf{q})$ frequencies are obtained from equations (\ref{acoust}) and (\ref{optic}), where, in place of $c_{s}(\mathbf{q})$, the generalized $c_0 (\mathbf{q})$ function is used and values for $\omega_{0}$ and $Y_0$ have been fixed. Note that, for the considered cubic lattices, the conventional cell with lattice parameter $a$ contains $n=1$ cage (simple cubic) or $n=2$ cages (body centered cubic), which implies $V= N a^3 /n$. To express the specific heat in $J/(K\cdot mol)$ unit, the factor preceding the sum in equation (\ref{SpecHeat}) has to be replaced with $3R/N$, where $R$ is the gas constant factor.
This approach was shown to be effective in the description of the phonons specific heat of LaB$_6$ and was subsequently used in the analysis of CeB$_6$ magnetic entropy \cite{Amara2020}. \newline

 \begin{figure}
\begin{center}
\includegraphics[width=12cm]{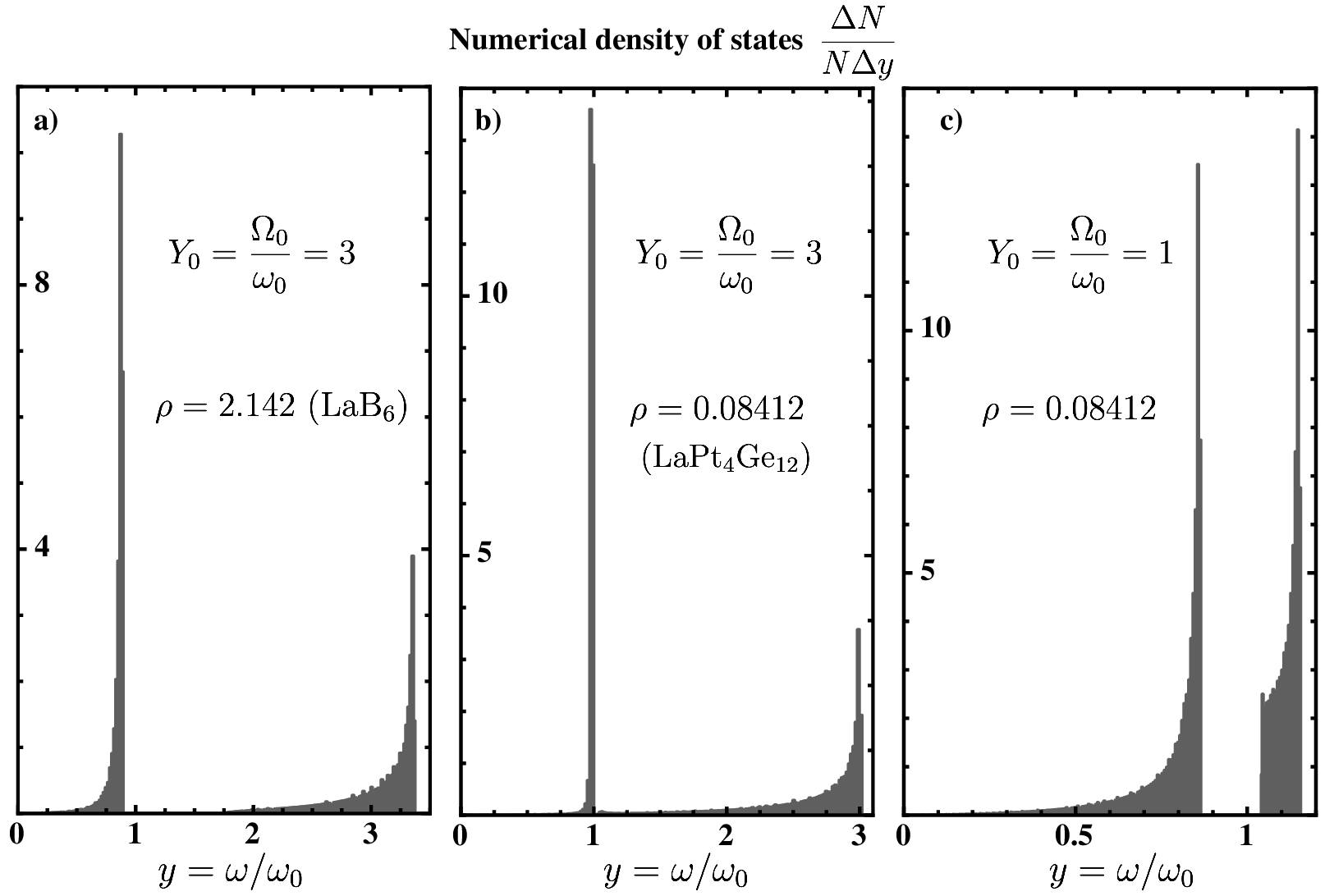}
\caption{\label{DensEtats} Numerical phonon densities of states as functions of the reduced frequency $y=\omega / \omega_0$,  for the same examples as in figure \ref{disp}. $\mathbf{a)}$ Heavy guest ($\rho >1$), hard lattice ($Y_0 = 3$) case, the $\rho$ mass ratio of LaB$_6$ and a simple cubic lattice of cages. $\mathbf{b)}$ Light guest, hard lattice case for the mass ratio of LaPt$_{4}$Ge$_{12}$, $Y_0 = 3$ and a body centered lattice.  $\mathbf{c)}$ Light guest, softer lattice ($Y_0 = 1$), for the mass ratio and body centered lattice of LaPt$_{4}$Ge$_{12}$.}
\end{center}
\end{figure}

Using the set of $N$ evenly distributed samples in the representative volume, a spectral analysis can be performed: i.e. extracting from this set the distribution in energy of the phonons. This involves counting the number of samples $\Delta N$ inside energy intervals of fixed width $\Delta y$, where $y$ is the energy in $\omega_0$ unit. If $N$ is large enough, this discrete approximation provides a good representation of what the continuous distribution would look like. Figure \ref{DensEtats} shows these numerical energy distributions, for an energy range split into a 100 intervals, for the three dispersion cases shown in Figure \ref{disp}: \newline
\begin{itemize}
\item [-] case $a)$, for a set of $\approx 6\cdot 10^5$ samples in the representative volume. Inspired by the LaB$_6$ example of heavy guest atoms inside a rigid lattice of cages, shows features of both Debye and Einstein models.  The increasing density of acoustic phonons, typical of the Debye model, is followed by a flattening of the acoustic branch, which results in a distinct Einstein-like peak. The heavy guests results in a large separation with the optical branch, this latter having negligible influence at low temperature. 
\item [-] case $b)$, for $\approx 1.2 \cdot 10^6$ samples in the representative volume. Example based on the mass ratio of LaPt$_{4}$Ge$_{12}$ (i.e., in relative terms, a light guest atom) and considering a rigid lattice of cages. This example is the closest to the Einstein approximation, with a huge peak dominating the distribution. This features results from a small gap, with the acoustic and optical branches flattening at similar energies.
\item [-] case $c)$, again for $\approx 1.2 \cdot 10^6$ samples in the representative volume. Example with the same mass ratio of LaPt$_{4}$Ge$_{12}$, but considering a softer lattice of cages. It exhibits a more pronounced Debye-like distribution. However, due to the small gap and soft lattice, the flattened optical branch also contributes to the specific heat at intermediate temperatures. It is worth noting that this type of distribution resembles spectra obtained from inelastic neutron scattering on polycrystalline LaPt$_{4}$Ge$_{12}$. In ref. \cite{Galera2015}, the low-energy region of the spectra shows a first peak at approximately 7.5 meV, then a second peak at 12.5 meV. This second peak is preceded by a shoulder around 10.5 meV, which matches the shape of the optical branch density for case $c)$.
\end{itemize}

\subsection{Phonon pressure}
The same two branches model should apply for computing the low temperature phonon contribution to the thermal expansion. The approach here used is the quasi-harmonic one, were the harmonic treatment of the phonons is corrected for slight variations of the phonons frequency as function  of the volume (see, for instance, ref. \cite{Ashcroft76}). One starts from the phonon contribution to the internal energy for a system of volume $V$, equivalent to $N$ reciprocal lattice points:
\begin{equation}
\label{Uph}
U_{ph} (T,V) = \frac{3}{2} \sum_{i=1}^N  \sum_{\mu =(a,o)} \hbar \omega_{\mu}(\mathbf{q}_i)+ 3 \sum_{i=1}^N  \sum_{\mu =(a,o)} \frac{\hbar \omega_{\mu}(\mathbf{q}_i)}{e^{ \frac{\hbar \omega_{\mu}(\mathbf{q}_i)}{{k_B}T}}-1}
\end{equation}

From the phonon internal energy, one has to derive the free energy : $F_{ph} (T,V) = U_{ph}(T,V) - T S_{ph}(T,V)$. This in order to introduce the phonon pressure :
\begin{equation}
\label{Pph}
P_{ph} (T,V) = - \left. \frac{\partial F_{ph}}{\partial V}  \right)_T  
\end{equation}

In absence of any internal degree of freedom for the cage guest (i.e. non-magnetic, in particular), one can assume that the $\omega_0$ frequency has negligible temperature dependence below room temperature. The same applies for the lattice of cages represented by the frequency $\Omega_0$. We shall consider that these two characteristic frequencies depend only on the volume $V$, which, in turn, also applies for the normal modes $\omega_{\mu}(\mathbf{q})$. In these conditions, the phonon pressure exerted on the crystal matrix can be expressed as in ref. \cite{Ashcroft76}:
\begin{equation}
\label{Pph}
P_{ph} (T,V) = P_0 - 3 \hbar \sum_{i=1}^N  \sum_{\mu =(a,o)}  \frac{d  \omega_{\mu}(\mathbf{q}_i)}{d V}  \frac{1}{e^{\frac{\hbar \omega_\mu\left(\mathbf{q}_i\right)}{k_B T}}-1}
\end{equation}
where $P_0$ is the zero temperature phonon pressure. For the small volume changes involved, the volume derivatives of the $ \omega_{\mu}(\mathbf{q})$ frequencies are considered as constants. 

\subsection{Phonons thermal expansion}
Introducing $\chi_0$,  the compressibility of the lattice, supposed constant in the temperature ranges of interest, the linear thermal expansion coefficient $\alpha_{ph}$ can be related to the phonon pressure:
\begin{equation}
\label{alphaph}
\alpha_{ph} = \frac{\chi_0}{3}  \left.  \frac{\partial P_{ph}}{\partial T} \right)_{V}
\end{equation}
\footnote{Note that, if one seeks a value for $\chi_0$, the cubic symmetrized elastic constants $C_{11}$ and $C_{12}$ are more easily found in the literature. We recall that $\chi_0$ can be written as:
\begin{equation*}
\chi_0 = \frac{3}{C_{11}+2 C_{12}}
\end{equation*}}
From equations \ref{Pph} and \ref{alphaph}, the thermal expansion coefficient can be expressed as:

 \begin{equation}
 \alpha_{ph} (T) = \frac{\chi_0}{3}  \sum_{i=1}^N  \sum_{\mu =(a,o)}\gamma_{\mu}(\mathbf{q}_i) \cdot c_{\mu}(\mathbf{q}_i , T)
\end{equation}

where, for each mode, are introduced:
\begin{itemize}
\item [-] an individual Gr{\"u}neisen parameter, 
\begin{equation}
\label{IndivGruen}
\gamma_{\mu}(\mathbf{q}_i) = - \frac{V}{\omega_{\mu}(\mathbf{q}_i)}\frac{d  \omega_{\mu}(\mathbf{q}_i)}{d V}
\end{equation}
\item [-] a contribution to the specific heat per unit volume:
\begin{equation}
\label{SpecHeatIndiv}
c_{\mu}(\mathbf{q}_i , T)=\frac{3 k_{\mathrm{B}}}{V} \left(\frac{\hbar \omega_{\mu}(\mathbf{q}_i)}{k_{\mathrm{B}} T}\right)^2 \frac{e^{-\frac{\hbar \omega_{\mu}(\mathbf{q}_i)}{k_{\mathrm{B}} T}}}{\left(1-e^{-\frac{\hbar \omega_{\mu}(\mathbf{q}_i)}{k_{\mathrm{B}} T}}\right)^2} 
\end{equation}
\end{itemize}

To detail the expression for $ \alpha_{ph}$, one needs to define the volume derivative for all $\omega_{\mu}(\mathbf{q}_i)$ frequencies. Using the generalized dispersion relations,  these derivatives can be related to the frequencies $\omega_0$ and $\Omega_0$. Indeed, for a given mode, the cosine $\cos (2 \pi {q}{d})$ from equation (\ref{EqDisp1}) is independent of the volume. With respect to an infinitesimal volume variation $dV$, the following differential, involving the mode of frequency $\omega_{\mathbf{q}}$, has to cancel:
\begin{equation}
d\left[ {\frac{{{ \omega_{\mathbf{q}}}^2}}{{{\Omega_0}^2}} + \rho \frac{{{ \omega_{\mathbf{q}}}^2}}{{{\Omega_0}^2}}\frac{{{\omega _0}^2}}{{{\omega _0}^2 - { \omega_{\mathbf{q}}}^2}}} \right] = 0
\end{equation}
This translates into a relation between the differentials of $\omega_{\mathbf{q}}$,  $\omega_0$ and $\Omega_0$. Introducing the peak function $f_{\rho}$ of the reduced frequency $y=\omega_{\mathbf{q}} / \omega_0$, defined for a given mass ratio $\rho$:
\begin{equation}
f_{\rho}(y) = \frac{\rho}{\left( 1 - y^2 \right)^2 + \rho }
\end{equation}
the partial derivatives of $\omega_{\mathbf{q}}$ read as:
\begin{equation}
{\left. {\frac{{\partial { \omega_{\mathbf{q}}}}}{{\partial {\omega _0}}}} \right)_{{\Omega_0}}}=f_{\rho}(\frac{\omega_{\mathbf{q}}}{\omega _0}) \frac{{\omega_{\mathbf{q}}}^3}{{\omega _0}^3}
\end{equation}
and 
\begin{equation}
{\left. {\frac{{\partial { \omega_{\mathbf{q}}}}}{{\partial {\Omega_0}}}} \right)_{{\omega _0}}} =\frac{ \omega_{\mathbf{q}}}{\Omega_0}\left( 1 - f_{\rho}(\frac{\omega_{\mathbf{q}}}{\omega _0}) \frac{{\omega_{\mathbf{q}}}^2}{{\omega _0}^2} \right)
\end{equation}
As changes in $\omega _0$ and $\Omega_0$ are uniquely ascribed to the volume change $dV$ :
\begin{equation}
d \omega_{\mathbf{q}} = f_{\rho}(\frac{\omega_{\mathbf{q}}}{\omega _0}) \frac{\omega_{\mathbf{q}}^3}{{\omega _0}^3}V \frac{d {\omega _0}}{dV} \frac{dV}{V} +\frac{ \omega_{\mathbf{q}}}{\Omega_0}\left( 1 - f_{\rho}(\frac{\omega_{\mathbf{q}}}{\omega _0}) \frac{{\omega_{\mathbf{q}}}^2}{{\omega _0}^2} \right)V \frac{d \Omega_0}{dV} \frac{dV}{V}
\end{equation}
One can then express the individual Gr\"uneisen parameter of equation (\ref{IndivGruen}) as:

\begin{equation}
\gamma_{\mu}(\mathbf{q}_i) = f_{\rho}(\frac{\omega_\mu\left(\mathbf{q}_i\right)}{\omega _0}) \frac{\omega_\mu\left(\mathbf{q}_i\right)^2}{{\omega _0}^2}\gamma _0 +\left( 1 - f_{\rho}(\frac{\omega_\mu\left(\mathbf{q}_i\right)}{\omega _0}) \frac{\omega_\mu\left(\mathbf{q}_i\right)^2}{{\omega _0}^2} \right)\Gamma _0
\end{equation}

where a Gr\"uneisen parameter is introduced for each characteristic frequency,
\begin{equation}
\gamma _0 = - \frac{V}{\omega _0} \frac{d{\omega _0}}{dV} 
\end{equation}
for the natural rattler frequency $\omega _0$ and :
\begin{equation}
\Gamma _0 = - \frac{V}{\Omega_0} \frac{d{\Omega_0}}{dV} 
\end{equation}
for the averaged top frequency $\Gamma _0$ of a lattice of empty cages.\newline
Thus, the expression for the thermal expansion linear coefficient:

\begin{equation}
 \alpha_{ph} (T) = \frac{\chi_0}{3}  \sum_{i=1}^N  \sum_{\mu =(a,o)} c_{\mu}(\mathbf{q}_i , T) \Big\{ \Gamma _0 + (\gamma_0 -\Gamma _0 )f_{\rho}(\frac{\omega_\mu\left(\mathbf{q}_i\right)}{\omega _0}) \frac{\omega_\mu\left(\mathbf{q}_i\right)^2}{{\omega _0}^2} \Big\} 
\end{equation}

In the expression above, one can identify the phonon contribution to the sample specific-heat $C_{ph}$ and its corresponding Gr\"uneisen contribution to the thermal expansion. There is an additional term, specific to the cage phonons model, proportional to the function $D_{ph}$ of the temperature, also dimensionally a specific heat per unit volume:
\begin{equation}
D_{ph} (T) =  \sum_{i=1}^N  \sum_{\mu =(a,o)} f_{\rho}(\frac{{\omega_{\mu}(\mathbf{q}_i)}}{\omega _0}) \frac{{\omega_{\mu}(\mathbf{q}_i)}^2}{{\omega _0}^2} c_{\mu}(\mathbf{q}_i , T)
\end{equation}

\begin{figure}
\begin{center}
\includegraphics[width=12cm]{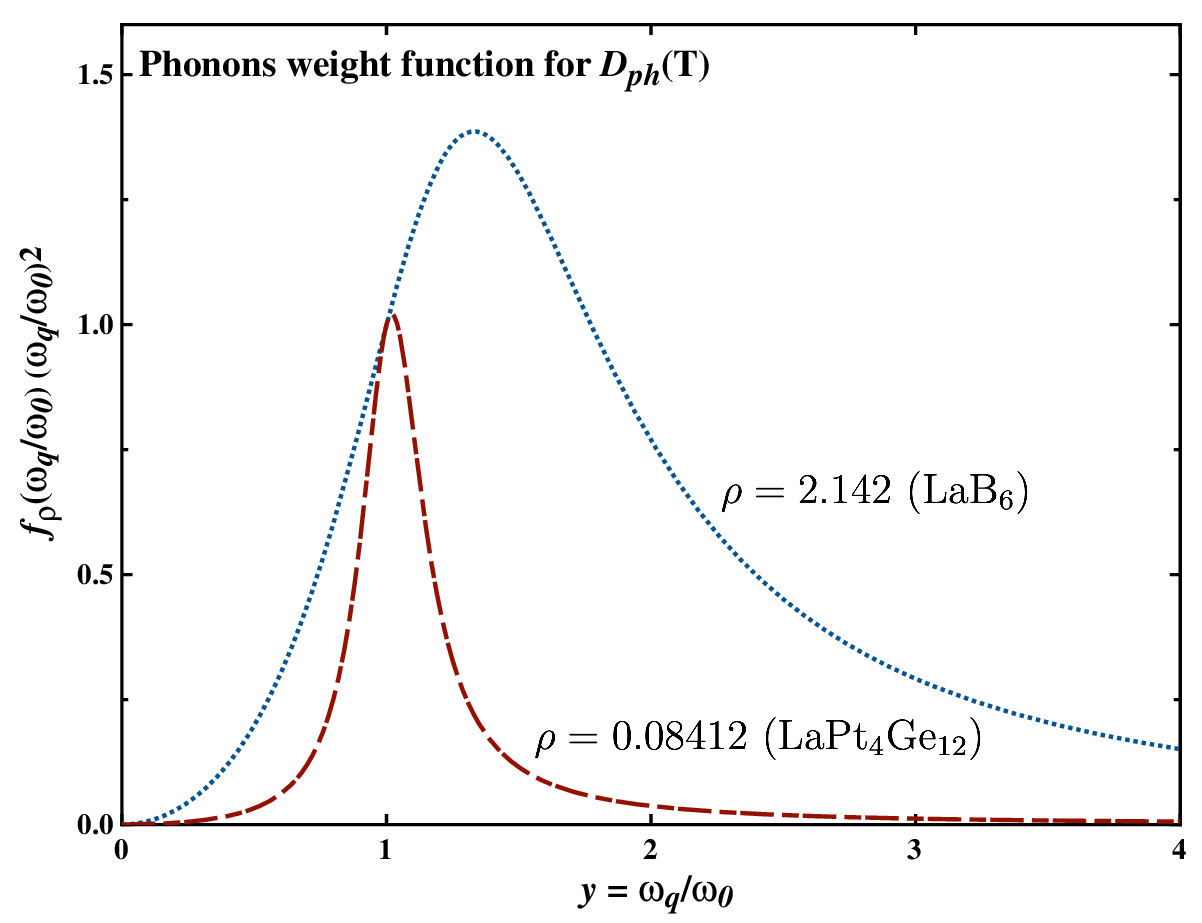}
\caption{\label{dilatweight} Plots of the weight functions $f_{\rho}(\frac{\omega_{\mathbf{q}}}{\omega _0}) \frac{{\omega_{\mathbf{q}}}^2}{{\omega _0}^2}$, for LaB$_6$ and LaPt$_4$Ge$_{12}$, active in the $D_{ph}$ term of the thermal expansion.  These two examples are drastically different in terms of the mass ratio $\rho$, which defines the width of the $f_{\rho}$ function, with a heavy guest for LaB$_6$ against a heavy cage for LaPt$_4$Ge$_{12}$.}
\end{center}
\end{figure}

As shown in figure \ref{dilatweight}, the weight function $f_{\rho}(\frac{\omega_{\mathbf{q}}}{\omega _0}) \frac{{\omega_{\mathbf{q}}}^2}{{\omega _0}^2}$ increases the influence of modes with frequencies close to $\omega _0$. In approximative terms, $D_{ph}$ can be considered as the specific heat from modes belonging to the flattened phonon branches around $\omega _0$ (i.e., similar to a Einstein model specific heat). 
The expression for the phonons thermal expansion coefficient then reads as:
\begin{equation}
\label{alphaPh}
\alpha_{ph}(T) = \frac{\chi_0\;\Gamma _0}{3} \Big\{ C_{ph}(T)+\left( 1- \frac{\gamma _0}{\Gamma _0} \right)  D_{ph}(T) \Big\} 
\end{equation}

where $C_{ph}$ is the phonons specific heat and $D_{ph}$ the just above discussed flattened branches 'specific heat'.  Equation (\ref{alphaPh}) shows that the phonons' model we use is not consistent with the Gr\"uneisen rule. This is no surprise, as the model is based on two frequency parameters, $\omega_0$ and $\Omega_0$. As soon as these frequencies differ in their volume relative derivatives (i.e. $\gamma _0 \neq \Gamma _0$), $D_{ph}$ interferes in the 'proportionality' between $\alpha_{ph}$ and $C_{ph}$.   Instead of a constant parameter, one can introduce an effective Gr\"uneisen function $\gamma_{eff}(T)$ that reads as :
\begin{equation}
\label{Gruenf}
\gamma_{eff}(T) = \Gamma _0\Big\{ 1+\left( 1- \frac{\gamma _0}{\Gamma _0} \right) \frac{ D_{ph}(T)}{C_{ph}(T)} \Big\} 
\end{equation}

 \begin{figure}
\begin{center}
\includegraphics[width=12cm]{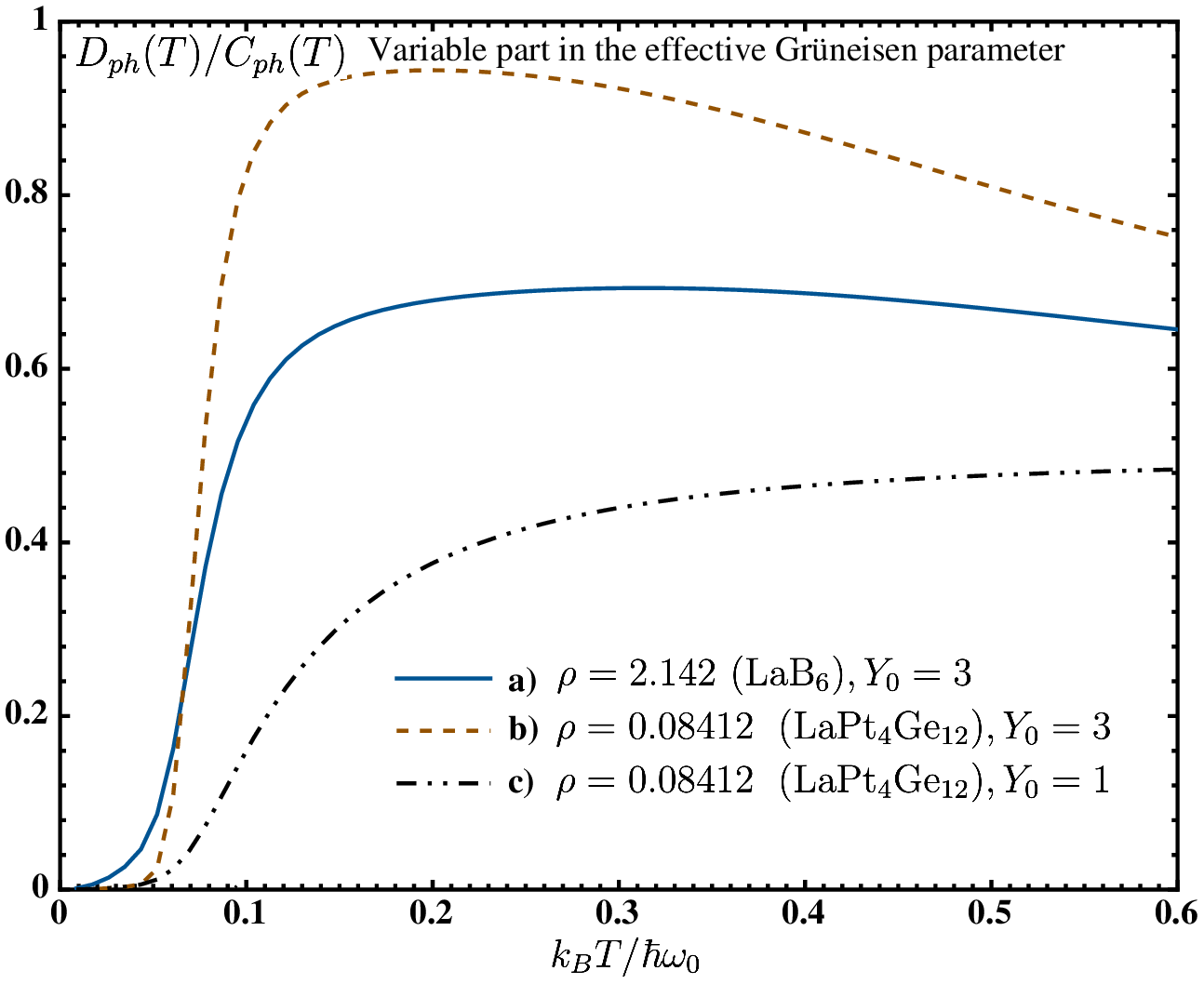}
\caption{\label{Vargamma} Plots of the ratios $ D_{ph}(T)/C_{ph}(T)$, as functions of the reduced temperature $k_{\mathrm{B}} T / \hbar \omega_0 $, for the same a), b) and c) examples as in figures \ref{disp} and \ref{DensEtats}. These ratios define the variable part in the $\gamma_{eff}(T)$ 'Gr\"uneisen' function, i.e. the deviation from the Gr\"uneisen rule. }
\end{center}
\end{figure}
As the physical contexts underlying the two types of vibrations, those of the guest inside the cage and those of the cage within its lattice, are quite distinct, $\gamma _0$ and $\Gamma _0$, are unlikely to be identical. The latter case is more similar to simple lattices discussed in the literature, where typical values for metals approach 2 \cite{White65}. As regards ${\gamma _0}$, the crudest, non-harmonic, approach is that of a box potential that would yield ${\gamma _0}= \frac{2}{3}$.  Thus, there is no a priori reason to ignore the contribution of the ratio $D_{ph}(T) / C_{ph}(T)$ to $\gamma_{eff}(T)$. This ratio and its temperature dependence depend only on $\omega_0$, $\Omega_0$ and the mass ratio $\rho$. It can be computed for the examples of figures \ref{disp} and \ref{DensEtats}. The curves on figure \ref{Vargamma} show two distinct regimes, the low-temperature Gr\"uneisen rule breaking down at temperatures as low as $T \approx 0.05\;  \hbar \omega_0 / k_{\mathrm{B}} T $.  In this low temperature regime the Gr\"uneisen parameter identifies with $\Gamma _0$: the acoustic branch of the lattice of cages is dominant in the phonons' thermal expansion, drawing a parallel with the Debye approximation for the specific heat.  Above $ T \approx 0.1\;  \hbar \omega_0 / k_{\mathrm{B}} T $, the influence of the Einstein-like flattened phonon branches, related to the rare-earth vibration, becomes evident: after a steep increase, the $D_{ph}(T) / C_{ph}(T)$ ratio stabilizes, leading to a new, nearly constant Gr\"uneisen 'parameter' at intermediate temperatures.  This plateau in $D_{ph}(T) / C_{ph}(T)$ is more pronounced when there is a large gap between the two phonon branches (example $a$, with the mass ratio of LaB$_6$). Since it depends on the unknown contrast between $\gamma _0$ and $\Gamma _0$, at this stage, the sign and amplitude of the step in $\gamma_{eff}(T)$ cannot be predicted.

\section{Experiments}
\subsection{Samples}
The LaB$_6$ samples used are single crystals grown at the Frantsevich Institute in Kyiv. The synthesis process is detailed in the supplementary material of reference \cite{Azarevich2024}. From the zone melting processed rod, a single crystalline platelet was cut with the larger face perpendicular to a twofold axis. This sample was used for the thermal expansion measurements, its diagonal of 6,02 mm length, also along a twofold direction, being the one sensed in the apparatus. For the specific heat measurement, a much smaller sample from the same batch was used, with a mass $ m = 6$ mg. \newline
The polycrystalline sample of LaPt$_{4}$Ge$_{12}$ was prepared by first melting stoichiometric amounts of the three elements in a cold crucible under an argon atmosphere. Then, to complete the skutterudite phase formation, the ingot was sealed in an evacuated quartz tube and annealed at 780$^{\circ}$ C for 10 days.  Powder X-ray diffraction revealed that the annealed sample was free of parasitic phases, with the X-ray pattern consistent with the filled skutterudite phase. A piece of this ingot, with a sensed length of 5.98 mm, was used for thermal expansion measurements. From the same ingot, a much smaller piece weighing m = 6.05 mg was detached for the specific heat measurements.

\begin{figure}
\begin{center}
\includegraphics[width=12cm]{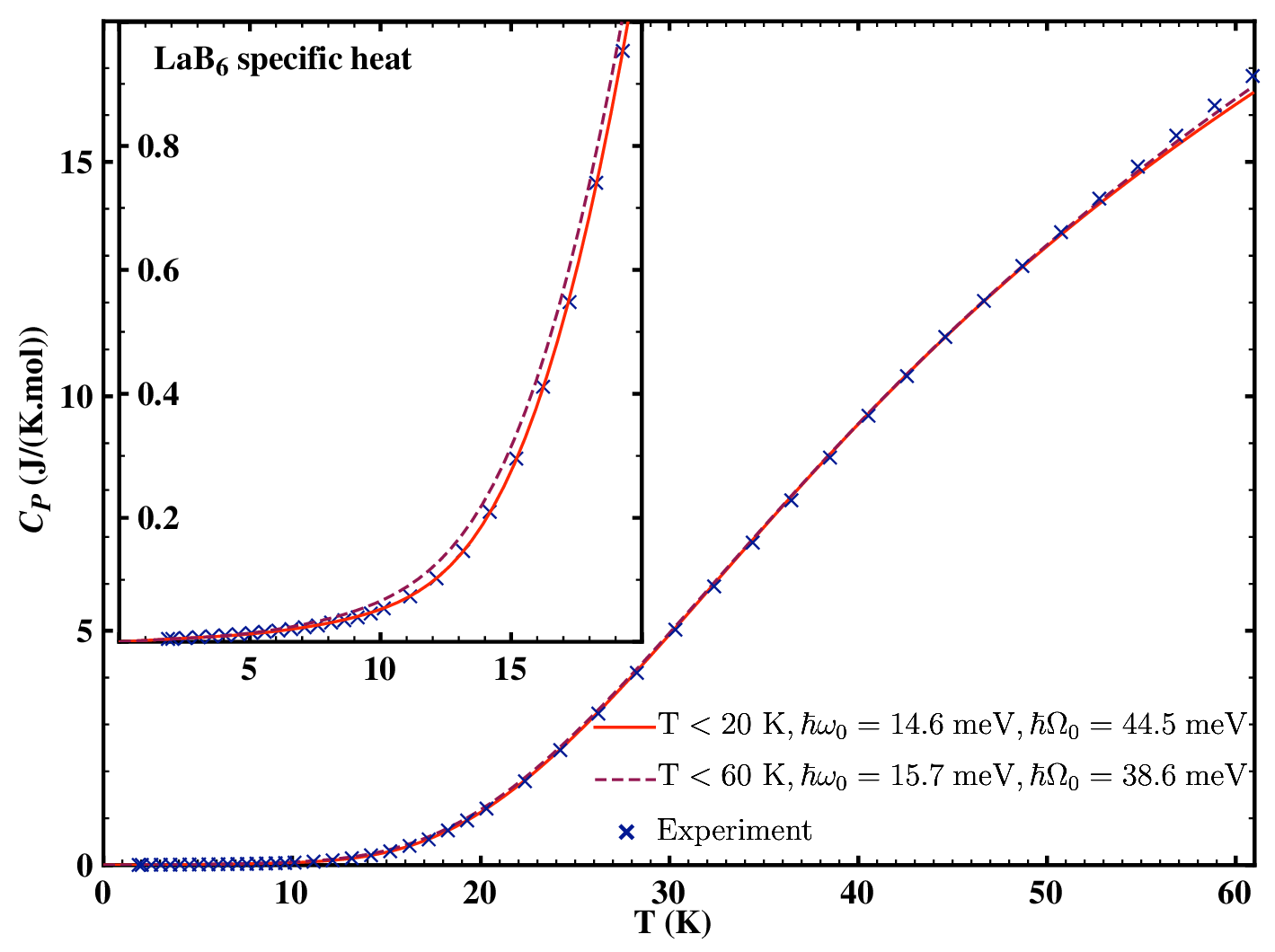}
\caption{\label{CpLaB6} Specific heat data for LaB$_6$ (crosses). The inset details the low temperature range. The lines are least squares fits using the two frequencies $\omega_0$ and $\Omega_0$ model.  The continuous line is a fit to the experimental data for temperatures below 20 K, whereas the doted line is a fit to the full set of data, up to 60 K. The respective, optimized values for $\omega_0$ and $\Omega_0$, are displayed in the legend. The value $\gamma_e = 2.3$ mJ/(K$^2$.mol) for the electronic contribution coefficient to the specific heat is not adjusted (see text).  }
\end{center}
\end{figure}

\subsection{Specific heat measurements}
\begin{figure}
\begin{center}
\includegraphics[width=12cm]{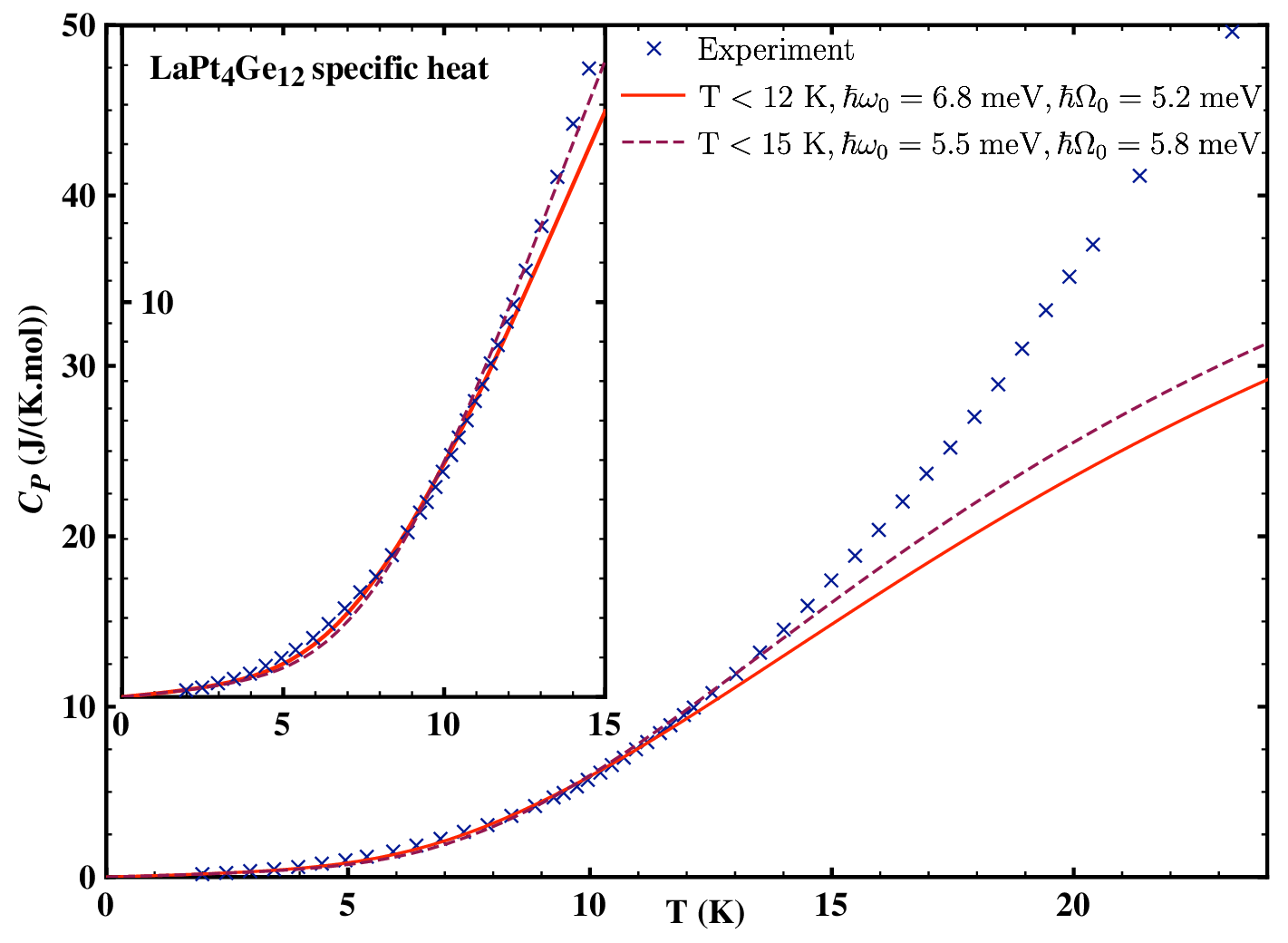}
\caption{\label{CpLaPtGe} Specific heat measurements for LaPt$_{4}$Ge$_{12}$ under a 1.2 T applied magnetic field, in order to suppress the superconducting transition at $T_C$ = 8.3 K, with low temperature detail in the inset. The lines are least squares fits using the two frequencies $\omega_0$ and $\Omega_0$ model and fixed value $\gamma_e = 80$ mJ/(K$^2$.mol) coefficient for the electronic contribution to the specific heat.  The continuous line is a fit to the experimental data for temperatures below 12 K while the doted line extends the range to 15 K. The respective optimized values for $\omega_0$ and $\Omega_0$ are displayed in the legend. }
\end{center}
\end{figure}

The specific heat measurements were performed using the relaxation technique in an automated\cite{Hwang1997} commercial Quantum Design PPMS system. The thermal coupling between the sample and the setup platform was improved by use of Apiezon N grease, which is accounted for thanks to the addenda measurements. For both the LaB$_6$ and LaPt$_{4}$Ge$_{12}$ sample, the default two time constants fitting of the relaxation process, was used. In order to avoid the anomaly resulting from the superconducting transition in LaPt$_{4}$Ge$_{12}$ at $T_C = 8.3$ K \cite{Gumeniuk2008}, the specific heat was measured under a 1.2 T applied magnetic field. The measurements for LaB$_6$ and LaPt$_{4}$Ge$_{12}$ are respectively shown in figures \ref{CpLaB6} and \ref{CpLaPtGe}.

\subsection{Thermal expansion measurements}
\subsubsection{Use of a magnetostriction setup}
The thermal expansion was measured using the custom-built magnetostriction setup of the Institut Néel. This system consists of a capacitance cell machined from a copper-beryllium alloy, housed in a helium flux cryostat equipped with a superconducting magnet that provides a horizontal magnetic field up to 6 T. The cell can be rotated along a vertical axis, allowing for length measurements both along and perpendicular to the applied magnetic field, within a temperature range of 2 to 300 K. The sample is glued on a Cu-Be sample holder with Stycast 2850FT, the other end being free but elastically pressed against the moving electrode. The cell and associated capacitance bridge can detect changes in length as small as 1 \AA, enabling the system to sense relative expansions down to $10^{-7}$ for samples of a few millimeters in length. 
The main issue, in using this system for thermal expansion measurement, is that it was originally designed for field variations at constant temperature. For variable temperature measurements, corrections must be made for the cell's thermal hysteresis. This is done by measuring a reference sample, a sphere made of the same copper-beryllium alloy as the cell, following a temperature cycle identical in rate and interval to that applied to the investigated sample. 

\subsubsection{Linear thermal expansion coefficient}
\label{linthermexp}
\begin{figure}
\begin{center}
\includegraphics[width=12cm]{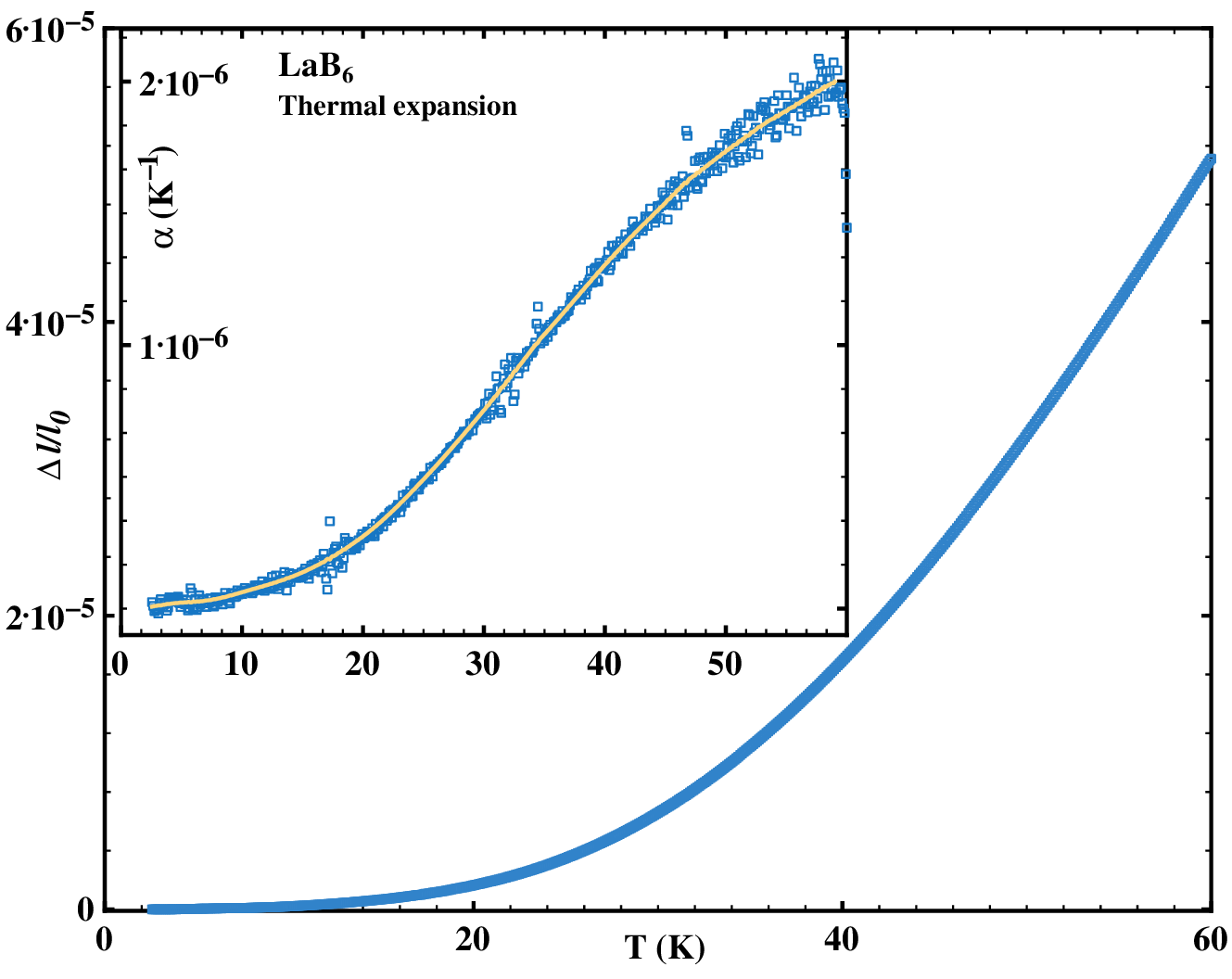}
\caption{\label{alphaLaB6} Thermal expansion of LaB$_6$, as the relative change in length with respect to $l_0$ at 2.5 K, measured on a single crystal with a capacitance dilatometer while cooling down from 60 K to 2.5 K. The inset gives the linear thermal expansion coefficient $\alpha$, as a numerical derivative (scattered points) and with applied low-pass filtering (light color, solid line, for a filtering constant $\tau=1.5$ K).}
\end{center}
\end{figure}

\begin{figure}
\begin{center}
\includegraphics[width=12cm]{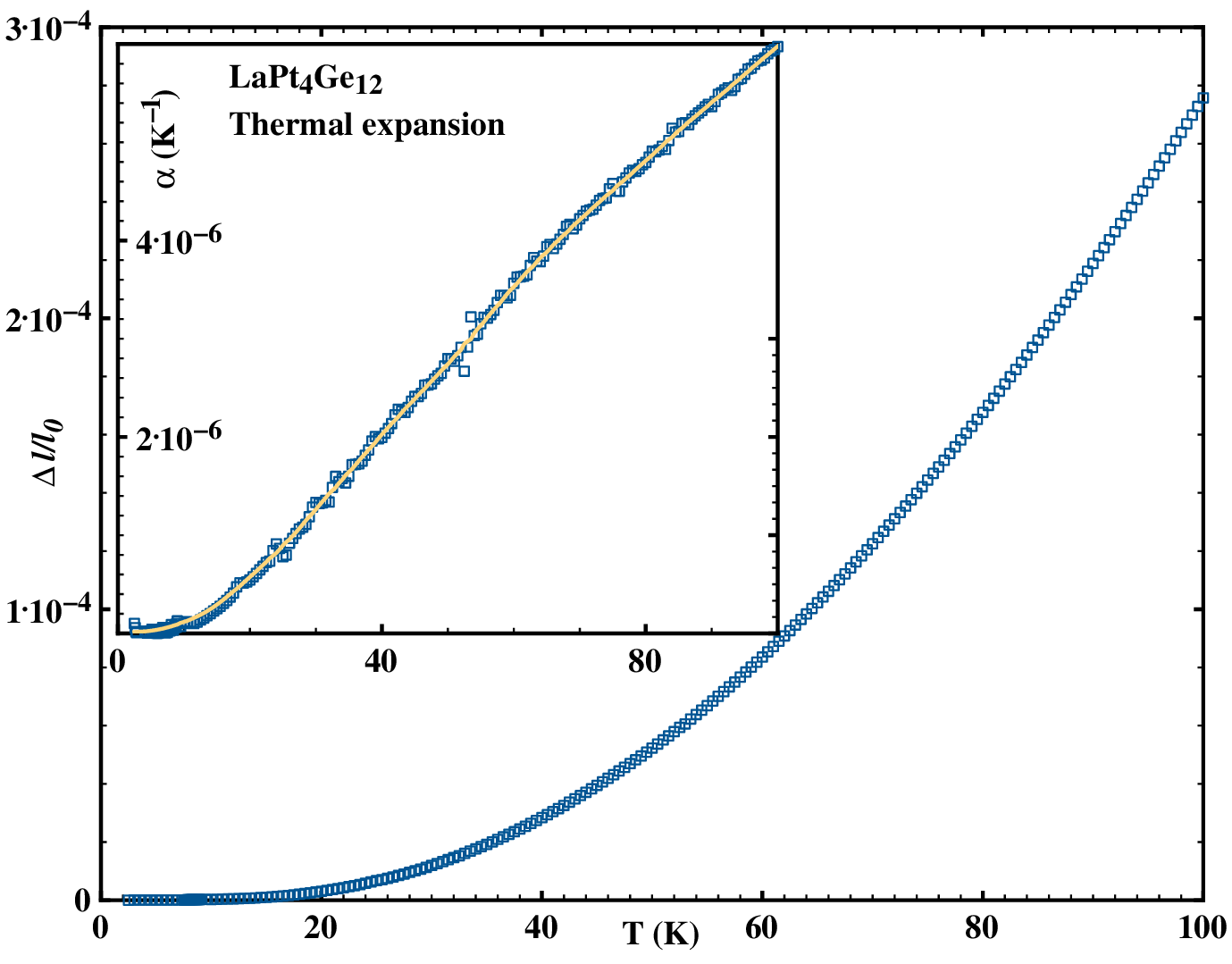}
\caption{\label{alphaLaPtGe} Thermal expansion of LaPt$_{4}$Ge$_{12}$, as the relative change in length with respect to the lowest temperature length $l_0$, measured on a polycrystalline sample with a capacitance dilatometer. The data below 9 K is measured under an applied magnetic field of 1.2 T to suppress the superconducting anomaly. Above 9 K, the measurements are for zero field. The inset gives the linear thermal expansion coefficient $\alpha$, as a numerical derivative (scattered open squares) and with applied low-pass filtering (solid line, filtering constant $\tau=3$ K).}
\end{center}
\end{figure}

To obtain the temperature variation of the linear thermal expansion coefficient $\alpha (T)$ from the sample length $l(T)$, one has to resort to a numerical approximation of the analytical relation :
\begin{equation}
\alpha(T) = \frac{1}{l} \left.  \frac{\partial l}{\partial T} \right)_{P}
\end{equation}
The linear thermal expansion coefficient, obtained from a numerical differentiation, is typically noisy when small temperature increments cause length changes at the limit of the apparatus sensitivity. An added difficulty is the occasional slip of the sample's contact point on the moving electrode, that yields parasitic peaks. To mitigate this dispersion in the data, a subsequent filtering can be applied. The filtering used here is a numerical analog to an RC low-pass filter, the input being the numerical derivative as function of the temperature. Instead of a time constant, a value for a temperature constant $\tau$ is visually adjusted in order to dampen the abrupt changes that develop on smaller temperature scales. Such a filtering introduces a delay in the signal with respect to the temperature evolution. This delay is essentially cancelled by successively applying the filtering both up and down the temperature list, the two resulting outputs being then averaged. As the principle is the numerical solving of a first-order differential equation, one has to carefully choose the initial values of the output in both directions. In our case, these initial values are obtained by minimizing a chi-squared reflecting the distance between the filtered output and the noisy input.  \newline
Figure \ref{alphaLaB6} shows the measurement relative change in length for LaB$_6$, as obtained under zero magnetic field and decreasing temperature between 60 K and 2.5 K at a rate of 0.2 K/min. In the inset, the linear thermal expansion coefficient is displayed, as obtained from simple direct numerical differentiation (scattered points) and using a low-pass filter for a constant $\tau=1.5$ K (continuous line). \newline
The measurements for LaPt$_{4}$Ge$_{12}$ are displayed on figure \ref{alphaLaPtGe}. The relative change in length $\Delta l/l$ is obtained from measurements at a small rate of 0.05 K/min from 2.5 K to 120 K, in zero magnetic field, in both cooling and heating conditions. The data from the two temperature variations were then averaged as an additional mean for compensating the cell hysteresis. As this paper focuses on the effect of the low energy phonons, the temperature range of the figure is restricted to values below 100 K. Also, LaPt$_{4}$Ge$_{12}$ undergoing a superconducting transition at $T_C = 8.3$ K, with a slight volume anomaly, the data below 9 K is replaced by measurements done under an applied magnetic field of 1.2 T. The inset of figure \ref{alphaLaPtGe} shows the linear thermal expansion coefficient temperature variation, from direct numerical differentiation and subsequent low-pass filtering (continuous line) with a filtering constant $\tau=3$ K .

\section{Analysis}
\subsection{Specific heat}
The two dominant contributions to the specific of a non-magnetic, metallic compound are:
\begin{itemize}
\item [-] the conduction electrons specific heat, here reduced to its linear, low temperature term : $C_e (T) = \gamma_e T$, with $\gamma_e$ a constant characteristic of the system.
\item [-] the phonon contribution : $C_{ph} (T)$  which, for calculation, requires some model for the phonon dispersion curves.
\end{itemize}
$C_{ph} (T)$ is here computed using the two frequencies, two branches, generalized dispersion curves for cage compounds. Using equation (\ref{SpecHeat}) and an appropriate sampling of the first Brillouin zone, the $C_{ph} (T)$ specific heat can be numerically calculated as a function of temperature for any given pair of frequency values $(\omega_0 , \Omega_0)$.
As regards the $\gamma_e$ electronic constant, it can be directly determined from the specific heat experimental data $ C_{p} (T)$. Here, the values used are obtained from the 0 K extrapolation of the plot of  $C_{p}/T$ as function of $T^2$. Once $\gamma_e$ is defined, values for $\omega_0$ and $\Omega_0$ can de retrieved from a least squares  fit to the experimental data.

\subsubsection{LaB$_6$}
Figure \ref{CpLaB6} shows, as functions of the temperature, the results of calculations superimposed with the specific heat data. The computed values include an electronic contribution for $\gamma_e = 2.3\pm0.1$ mJ/(K$^2$.mol), deduced from the plot of $C_P /T$ against $T^2$. The mass ratio $\rho=2.142$ corresponds to a lanthanum ion inside a cage of natural boron. To expedite the calculation of $C_{ph}$ (equation (\ref{SpecHeat})) during the fitting process, the number of samples in the representative volume of the first Brillouin zone of the cubic lattice was limited to 364, which is sufficient to approach the large-number limit, without noticeable effects at the graph's scale. The choice of the temperature range, used for the optimization of the  $\omega_0$ and $\Omega_0$ values, is crucial. Clearly, the model, based on only two phonon branches, treating the boron cage as a single object, cannot hold at higher temperatures, where it would violate the \emph{Dulong et Petit } law.  In figure \ref{CpLaB6}, two least squares refinements are displayed, one restricted to data for temperatures below 20 K (full line), the other generalized to the full set, up to 60 K (dashed line). It is apparent that, in both cases, the model struggles to describe the experiment above 50 K. For the extended set of temperatures, the description above 50 K is marginally improved at the expense of accuracy in the low temperature range (see the inset).\newline
The best fit below 50 K is obtained for the low temperature data set (T $<$ 20 K), which aligns with the model's expected limitations at higher temperatures. The refined values are $\hbar \omega_0=14.6$ meV and $\hbar \Omega_0=44.5$ meV, resulting in a ratio $Y_0 = \Omega_0 / \omega_0 =3.05$ close to the value 3 used in the $a)$ example in figures \ref{disp}, \ref{DensEtats} and \ref{Vargamma}. The value for $\hbar \omega_0$ slightly exceeds the inelastic neutron scattering \cite{Smith1985,Amara2020} investigation, which points to a value of approximately 13.5 meV. From this LaB$_6$ example, it can be concluded that the model accurately describes the phonon specific heat up to one-third of $\hbar \omega_0 / k_B$, i.e. about 50 K. Above this limit, higher energy phonon branches become influential, increasing the specific heat relative to the two frequencies model. In rare-earth hexaborides, at least, this limitation is inconsequential, as the relevant magnetic phenomena typically occur below 30 K. For instance, in CeB$_6$, this phonon description is effective in defining the temperature dependence of the magnetic entropy below 50 K\cite{Amara2020}.   

\subsubsection{LaPt$_{4}$Ge$_{12}$}
The preliminary analysis, from the plot of $C_P /T$ as function of $T^2$, yields a value for the electronic contribution to the specific heat: $\gamma_e = 80\pm6$ mJ/(K$^2$.mol), consistent with earlier determinations \cite{Gumeniuk2008}. The massive cage results in a small mass ratio $\rho=0.08412$. Subsequently, one can proceed with the fit of the experimental data by adjusting the values of $\omega_0$ and $\Omega_0$, using equation (\ref{SpecHeat}). The calculations in figure \ref{CpLaPtGe} are performed for 650 samples within the representative volume of the body centered first Brillouin zone. The fitting procedure is applied to data from reduced low-temperature intervals, either below 15 K (dotted line) or below 12 K (solid line). Indeed, to obtain a reasonable fit, it is necessary to significantly lower the upper temperature limit. It appears that low energy modes, unaccounted for in our two-frequency description, become influential at temperatures as low as 15 K, causing the experimental data to rapidly exceed the calculated values.  The fits shown in figure \ref{CpLaPtGe} yield similar values for $\omega_0$ and $\Omega_0$ (i.e. $Y_0 = \Omega_0 / \omega_0 \approx 1$), suggesting a flattened optical branch with a high density of modes just above $\omega_0$ (see the dispersion curves of example $c$ in Figures \ref{disp} and \ref{DensEtats}). However, when confronted with the inelastic neutron scattering results \cite{Galera2015}, the fitted values are not satisfactory. The lowest energy phonon peak in the spectra is observed around 7.5 meV, while the specific heat fits in figure \ref{CpLaPtGe} would correspond to a peak slightly above 5 meV. The same inelastic scattering data reveals the proximity of phonon branches, unaccounted in the model, with the closest peak at about 15.5 meV. Below 20 K, the fit compensates for these missing contributions by lowering the energies of the two included branches.

\subsection{Testing the Gr\"uneisen rule}
An experimental assessment of the Gr\"uneisen rule requires, over a given temperature range, to define the ratio between the thermal expansion coefficient, $\alpha (T)$, and the specific heat $C_{P} (T)$. Measurements of these physical quantities for metals include contributions from the conduction electrons, whereas the present work focuses on the phonon contributions only. Since the conduction electrons terms, both for the thermal expansion and specific heat, are expected to vary linearly with the temperature across a broad range, more relevant quantities for an illustrative graph are $\alpha (T)/T$ and $C_{P} (T)/T$. The electronic contributions are thus reduced to constants and visually neutralized in the temperature evolution. In the following, the analysed plots are those of $\alpha (T)/T$ as function of $C_{P} (T)/T$.  If the Gr\"uneisen function $\gamma_{eff}$, introduced in equation (\ref{Gruenf}), is relatively constant over an extended temperature range, the graph will display a corresponding strait line. The slope of this line, that averages $\gamma_{eff}$, identifies with a local Gr\"uneisen parameter. Such plots require, at each temperature, experimental determinations for both $\alpha (T)$ and $C_{P} (T)$. Since the thermal expansion and specific heat data were collected separately, for different sets of temperatures, an extrapolation is needed. For each temperature in the specific heat dataset, a corresponding $\alpha$ value is identified using the low-pass filtered $\alpha$ curve (see paragraph \ref{linthermexp}). Additionally, the specific heat unit is converted from $J.K^{-1}.mol^{-1}$ to $J.K^{-1}.m^{-3}$, so that a slope in the graph has the dimension of a compressibility, i.e. the inverse of a pressure.

\begin{figure}
\begin{center}
\includegraphics[width=12cm]{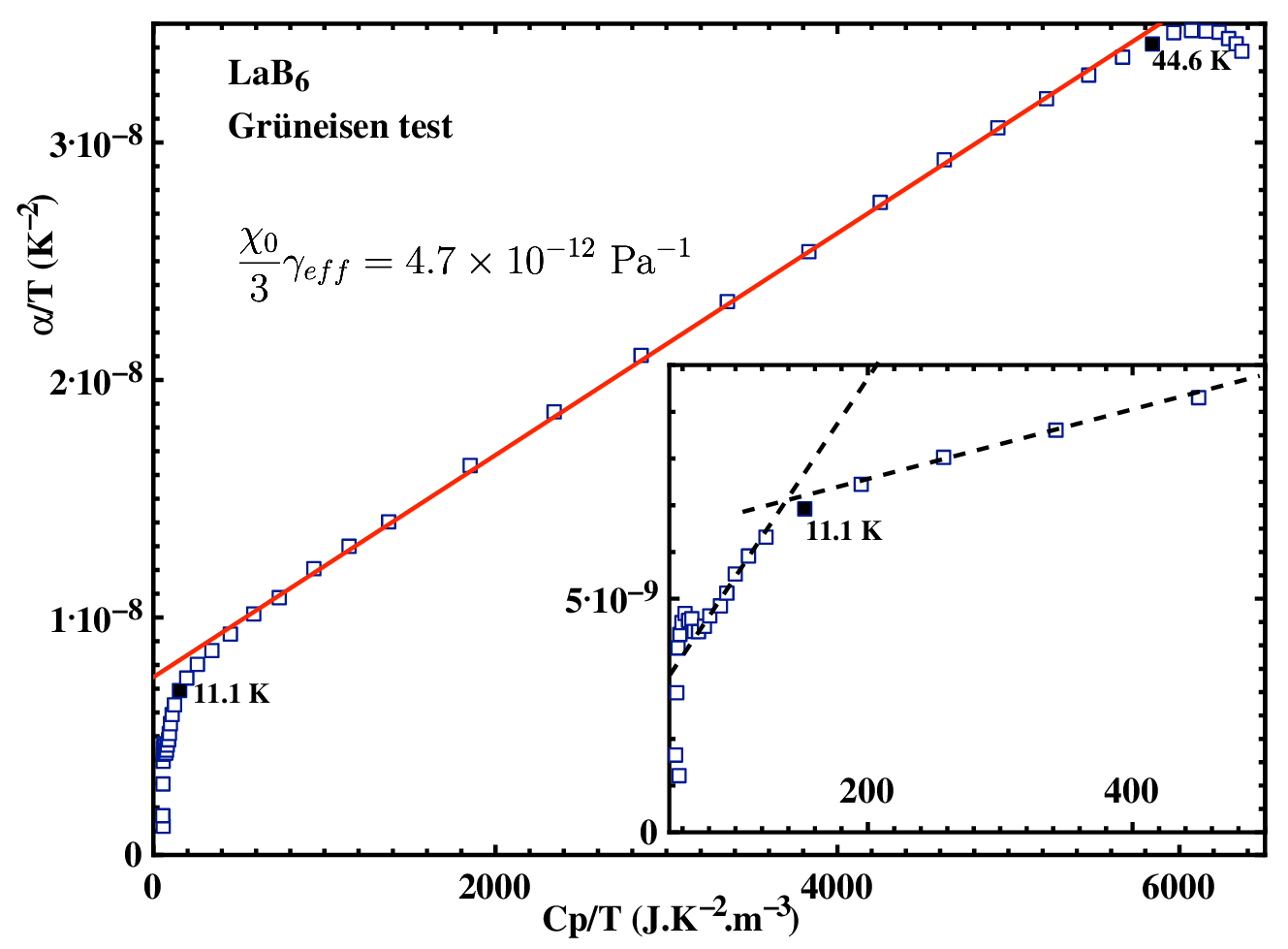}
\caption{\label{GrunAlphaCpLaB6} Test of the Gr\"uneisen rule for LaB$_6$ by plotting $\alpha /T$, where $\alpha$ is the linear thermal expansion coefficient for LaB$_6$, againts $C_p /T$, $C_p$ being the measured specific heat of LaB$_6$. For each temperature in the specific heat data a corresponding $\alpha$ value is identified using the low-pass filtered $\alpha$ curve (solid line in the inset of figure \ref{alphaLaB6}). Between 12 K and 45 K (see black squares), a nearly linear relation is observed that yields ${\chi_0 \gamma_{eff}}/{3}= 4.7\times 10^{-12}$ Pa$^{-1}$ for LaB$_6$. The inset highlights the low temperature region, with dashed lines pointing to a crossover at about 11 K. }
\end{center}
\end{figure}

\subsubsection{LaB$_6$}
Figure \ref{GrunAlphaCpLaB6} shows the test graph of the phonon Gr\"uneisen rule for LaB$_6$. Between 13 and 45 K, one observes a nearly linear relationship between $\alpha (T)/T$ and $C_{P} (T)/T$. According to equations (\ref{alphaPh}) and (\ref{Gruenf}), the slope of this linear region provides a value for the average of ${\chi_0 \gamma_{eff}(T)}/{3}$  in this temperature range. The compressibility ${\chi_0}$ at low temperature can be deduced from published elastic constants measurements  \cite{Nakamura1994}, yielding: $\chi_0 = 5.32\pm 0.06\times 10^{-12}$ Pa$^{-1}$. An estimate of $\gamma_{eff}$, the average of the Gr\"uneisen function over the linear region, is thus obtained as: $\gamma_{eff}=2.7$ .
\newline
Below 11 K, another regime is observed (see inset of figure \ref{GrunAlphaCpLaB6}), with a steeper slope, indicating a larger $\gamma_{eff}$ value. Although this region also appears fairly linear, caution is required as the dilatometer approaches its sensitivity limits below 8 K, resulting in increased relative scattering of $\alpha$ values, particularly below 5 K (see figure \ref{alphaLaB6}). Nevertheless, it can be concluded that $\gamma_{eff}(T)$ is significantly larger below 11 K than above. According to equation (\ref{Gruenf}), this implies that $\gamma_0$ is larger than $\Gamma_0$ (both should be positive), suggesting that the cage oscillator is more sensitive to volume changes than the lattice of cages. The crossover temperature between the two low-temperature regions is around $T_{co}=11$ K (see inset of figure \ref{GrunAlphaCpLaB6}). With parameters close to those of the specific heat fit, the $a)$ example in figure \ref{Vargamma} predicts a step in $\gamma_{eff}$ at $k_{\mathrm{B}} T_{co} / \hbar \omega_0 \approx 0.07$. This yields $\hbar \omega_0 \approx 13.5$ meV, a value in striking agreement with neutron spectroscopy and specific heat determinations.
\newline
Above 45 K, no Gr\"uneisen rule holds: additional phonon branches start to interfere, as observed in the specific heat data fits (figure \ref{CpLaB6}).

\begin{figure}
\begin{center}
\includegraphics[width=12cm]{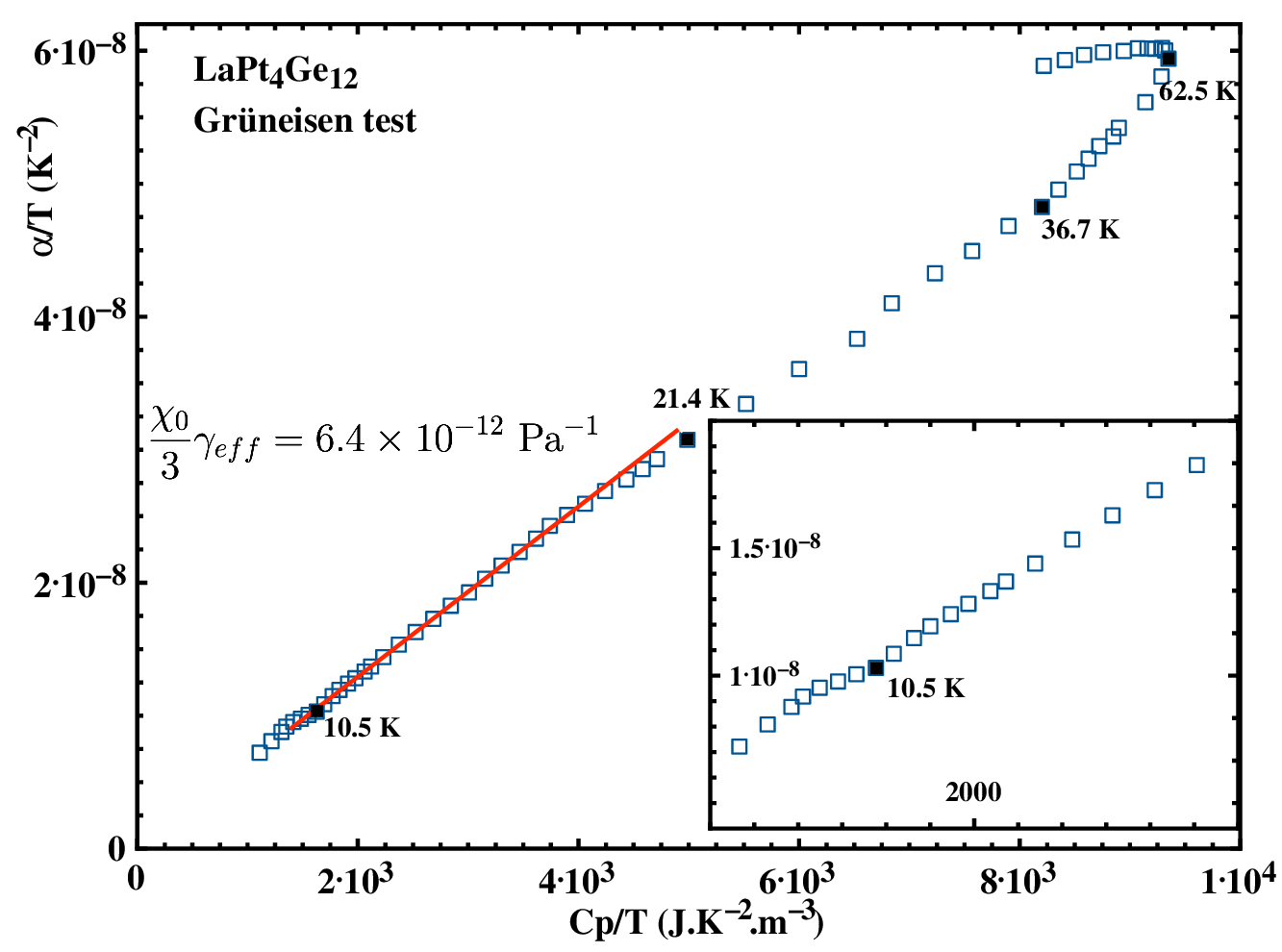}
\caption{\label{GrunAlphaCpLaPtGe} Test of the Gr\"uneisen rule for LaPt$_{4}$Ge$_{12}$ by plotting $\alpha /T$, where $\alpha$ is the linear thermal expansion coefficient for LaPt$_{4}$Ge$_{12}$, against $C_p /T$. For each temperature in the specific heat data set a corresponding $\alpha$ value is interpolated using the low-pass filtered data, limited to values above the superconducting transition. The inset highlights the low temperature range and the black squares mark particular temperatures. Between 11 K and 21 K, a linear relation is observed that yields ${\chi_0 \gamma_{eff}}/{3}= 6.4\times 10^{-12}$ Pa$^{-1}$ for LaPt$_{4}$Ge$_{12}$.}
\end{center}
\end{figure}

\subsubsection{LaPt$_{4}$Ge$_{12}$}
Figure \ref{GrunAlphaCpLaPtGe} illustrates the test of the Gr\"uneisen rule for phonons in LaPt$_{4}$Ge$_{12}$. The filtered thermal expansion data used for the plot were restricted to temperatures above the superconducting transition, to avoid artifacts resulting from the merging of different data sets (the data in figure \ref{alphaLaPtGe} are composite, combining zero-field measurements with data collected below 9 K under a magnetic field of 1.2 T). The graph reveals at least five distinct regions, separated by black squares. The two-frequency model can only account for the lower temperature range, where, between 11 and 21 K, an approximate Gr\"uneisen rule is observed. Within this range, the average ${\chi_0 \gamma_{eff}}/{3}$ is estimated to be $ 6.4\times 10^{-12}$ Pa$^{-1}$.  Due to the lack of available compressibility data for $\chi_0$, no value can be derived for the averaged Gr\"uneisen function $\gamma_{eff}$. In the lower temperature region, below the crossover at $T_{co}=10.5$ K (inset of figure \ref{GrunAlphaCpLaPtGe}), the slope decreases, which, unlike in the LaB$_6$ case, suggests a $\gamma_0$ value smaller than $\Gamma_0$, indicating a reduced influence of the cage oscillator on the thermal expansion in LaPt$_{4}$Ge$_{12}$. From the specific heat analysis, it was found that, among the treated dispersion examples, LaPt$_{4}$Ge$_{12}$ was closer to case $c)$, with $Y_0 \approx 1$. In figure \ref{Vargamma}, the step for case $c)$ is rather smooth, centered around $k_{\mathrm{B}} T_{co} / \hbar \omega_0 \approx 0.12$. Within the two-frequency model, this allows to estimate $\hbar \omega_0 \approx 7,5$ meV, a value larger than those derived from the specific heat peaks, but consistent with the neutron spectroscopy data. \newline
At higher temperatures, one can identify other quasi-linear regimes. It is tempting to speculate that these changes in the ${\chi_0 \gamma_{eff}}/{3}$ slope are due to the successive interference of higher energy flattened phonon branches. At about 62 K, there is a turning point in the plot, showing that both $C_{P}/T$ and $\alpha /T$ decrease above this temperature.

\section{Discussion}
This study clarifies the limits of the two-frequency model for rare-earth cage systems. In the case of LaB$_6$, the light and rigid lattice of boron cages isolates the two lowest phonon branches, allowing for an accurate description of the specific heat, up to temperatures equivalent to one-third of the natural frequency, $\omega_0$, of the Lanthanum guest. This is in contrast with the filled skuterrudite LaPt$_{4}$Ge$_{12}$. By analogy with LaB$_6$, one might expect an accurate specific heat description up to 30 K, yet discrepancies emerge as early as 15 K. This is due to the model's neglect of vibrations within the massive Pt$_{4}$Ge$_{12}$ cage, at frequencies close to that of the lanthanum guest atom, as revealed by the inelastic neutron scattering spectra. The two-frequency model may still be suitable for the thermodynamic analysis of other, lighter filled skutterudites, such as those in the RFe$_{4}$P$_{12}$ series.  Alternatively, one may try to extend the model, including additional phonon branches. In this process, the simple analytical forms for the dispersions would be lost, together with any advantage over a fully numerical approach, such as the Density Functional Theory. \newline
These same analytical expressions enable the extension of the two-frequency model to describe, within the quasi-harmonic approximation, the thermal expansion. This description involves two separate Grüneisen parameters, $\gamma_0$ and $\Gamma_0$, respectively related to the characteristic frequencies $\omega_0$ and $\Omega_0$.  Under these conditions, the system cannot conform to a simple Gr\"uneisen rule. Instead, the model predicts two successive Gr\"uneisen regimes: 
\begin{itemize}
\item [-] at low temperatures, an approximate Gr\"uneisen rule is followed, dominated by low-energy acoustic phonons, 
\item [-] at temperatures greater than one-tenth of $\hbar \omega_0 / k_{\mathrm{B}}$, the influence of the flattened phonon branches become significant, leading to a second approximate Gr\"uneisen rule.
\end{itemize}
In place of a single Gr\"uneisen parameter, it is then natural to define a Gr\"uneisen function of the temperature, as the ratio between the thermal expansion coefficient and the specific heat of phonons. This function is sensitive to changes in the distribution of vibrational modes across multiple phonon branches. By combining experimental data for specific heat and thermal expansion, appropriate plots can be generated to identify distinct temperature regions, highlighting changes in the Gr\"uneisen function. In the examples of LaB$_6$ and LaPt$_{4}$Ge$_{12}$, these plots confirm the expected regime transitions, moreover at temperatures that match the model's quantitative predictions.\newline
From an experimental standpoint, the main difficulty arrises from the limited sensitivity of our dilatometer. Below 10 K, changes in the sample length are difficult to detect, leading to large scatter in the experimental data. This hampers the graphical definition of the lower temperature regime, related to the cage lattice Gr\"uneisen parameter $\Gamma_0$. Consequently, a quantitative determination of $\Gamma_0$ and, subsequently, $\gamma_0$, is difficult. Furthermore, these determinations require complementary experimental data on low-temperature compressibility, which are lacking in the LaPt$_{4}$Ge$_{12}$ case.\newline
Nevertheless, the two-frequency model and its application to the investigation of the non-magnetic LaB$_6$ and LaPt$_{4}$Ge$_{12}$, provide a valuable insights into the thermal expansion mechanism in cage compounds. This understanding will help clarify the role of the orbital and spin degrees of freedom in systems where the lanthanum is replaced by a magnetic ion.
 
\section*{References}
\bibliography{/Applications/TeX/biblio}

\end{document}